# The Wave Function Cannot be a Real Wave – Then, Can We Speak of an Ontology of Particles?

*by Sofia D. Wechsler*

**Abstract**
Is the wave-function a physical reality traveling through our apparatus? Is it a real wave, or it is only a mathematical tool for calculating probabilities of results of measurements? Different interpretations of the quantum mechanics (QM) assume different answers to this question. It is shown in this article that the assumption that the wave-function is a real wave entails a contradiction with the predictions of the QM, when the special relativity is invoked. Therefore, this text concentrates on interpretations which conjecture that the reality that moves in our apparatuses is particles, and they move under the constraints of the wave-function.
The de Broglie-Bohm interpretation, which matches this picture, assumes that the particle travels along a continuous trajectory. However, the idea of continuous trajectories was proved to lead to a contradiction with the quantum predictions. Therefore this interpretation is not considered here.
S. Gao conjectured that the particle is in a permanent random and discontinuous motion (RDM). As it jumps all the time from place to place, the total set of occupied positions at a certain time is given by the absolute square of the wave-function. As motivation for his idea, Gao argued that if a charged particle were simultaneously in two or more locations at the same time, the copies of the particle would repel one another, destroying the wave-function. It is proved here that the quantum formalism renders this motivation wrong. Although refuting this motivation, the RDM interpretation is examined here. A couple of problems of this interpretation are examined and it is proved that they don't lead to any observable contradictions with the QM predictions, except one problem which seems to have no solution. In all, it appears that none of the wide-spread interpretations of the QM is free of contradictions.  .

**Abbreviations**
dBB     = de Broglie-Bohm
CSL     = continuous spontaneous localization
QM      = quantum mechanics
QO      = quantum object
PBR     = Pusey, Barrett and Rudolph
RDM     = random, discontinuous motion
w-f     = wave-function
w-p     = wave-packet

## 1. Introduction

The question whether the wave-function (w-f) is or is not a real wave traveling in our apparatus, has preoccupied for a long time the physicists. Pusey, Barrett and Rudolph (PBR) [1], [2], proved that the w-f must have a realistic counterpart which travels in our apparatus and which differs from one w-f to another. The most widespread interpretations of QM agree with this conclusion which seems almost trivial; indeed, in our apparatus it must travel some real entity, and it must differ from one w-f to a different one for producing different effects when tested.

- The de Broglie-Bohm (dBB) mechanics [3], [4], [5], assumes that the quantum object (QO) that travels in our apparatus consists of two items, a *particle* and a *wave*. The latter guides the trajectory of the particle by determining which trajectories are allowed and which are forbidden. However, the dBB mechanics conjectures that a particle cannot appear out of nothing and disappear into nothing. Therefore its trajectory must be continuous and not split. It was proved that the assumption of continuity of trajectories leads to a contradiction, [6]. Therefore this interpretation is not examined here.
- The 'consistent histories' interpretation [7] – [10] speaks in terms of *waves*. However, while claiming that it explains the measurement process without using the collapse postulate, in fact, in each history of a quantum system appears the collapse from a superposition of states to one of the members of the superposition, see examples in the chapters 12 and 13 of [8].
- The 'transactional' interpretation [11] – [14], suggests the existence of two types of *waves*. One is identical to the w-f, it is emitted by the source of the QO toward the detectors, and travels forward in time; the other type of waves, alien to the quantum formalism, is emitted by the detectors and travels backward in time. A 'hand-shake' is postulated to occur between the wave from the source and the wave from one of the detectors, and that determines which detector would click.
- Ghirardi, Rimini, Weber, Pearle, Gisin, and others, accepted the collapse of the w-f as a real phenomenon caused by a perturbing field [15] – [27]. The most advanced form of this proposal – the CSL model of collapse – is examined in detail in [27]. No clear claim is made whether the w-f is considered a real wave, but this interpretation does not speak of a particle. The perturbing field is supposed, in some works, to be the gravitational field [28] – [33]. However, this interpretation is problematic when applied to entanglements. For instance, if an experiment is performed dynamically, if the experimentalists change their minds and change the configuration of their apparatuses during the measurement (e.g. in the experiment with the photon singlet they change the orientations of the polarizers) the perturbing field should change according to their choices. No physical field changes according to human choices. D. Bedingham [34] tried to tailor a perturbing field to the polarization singlet, but he did not offer a solution to the mentioned problem.
- S. Gao [35] – [38], assumes that the w-f is a statistical entity, in the sense that the region occupied by the allowed positions of a particle at a given time determines the absolute square of the w-f at that time. The particle is supposed to be in a random discontinuous motion (RDM), i.e. jump from place to place. Thus, what has reality in Gao's interpretation of the QM is the particle, and the w-f is just a mathematical tool.

Only the Copenhagen interpretation of QM makes an exception [39]: it does not speak of a really existing item as PBR proved, be it wave or particle, it just doesn't deal with question of the w-f realism, but only with its predictions for tests results. The w-f is considered something that comprises all the information about the QO.

It is proved in the present article that the option that the w-f is a real wave, leads to a contradiction. Therefore, the idea that what travels in our apparatus is a particle, is examined. And since continuous trajectories are impossible, these particles should jump from a region of the space to another one, as conjectures the RDM interpretation. Gao motivated his interpretation of the QM as follows:

> "If the wave function represents a physical field, then it seems odd that there are (electromagnetic and gravitational) interactions between the fields of two electrons but no interactions between two parts of the field of an electron."[37]

The present text proves that this motivation is wrong, as the behavior of the QO is dictated by quantum laws, not by classical ones. But that does not invalidate the RDM interpretation. On the other hand it is shown here that this interpretation is not free of problems either, especially vis-à-vis the special relativity. In his work [35] Gao himself points to difficulties of the RDM. This text offers solutions for some of the difficulties, but has no solution for the problem with the special relativity.

**Note**: in this article the expression 'quantum particle' is frequently used. It may mean an elementary particle, or an atom or molecule for which the internal structure is ignored. In any case, the respective item is considered as described by the QM, not be the classical physics.

The rest of the sections are organized as follows: section 2 shows that the assumption of onticity of the w-f leads to a problem with the special relativity. Section 3 proves that the quantum formalism forbids interaction between two parts of the w-f of the same QO. Sections 4 and 5 try to apply the RDM to entanglements and shows that difficulties are encountered. For some of them, solutions are proposed, but there is a problem vis-à-vis the special relativity for which no solution can be suggested. Section 6 contains conclusions.

## 2. The assumption of onticity of the wave-function has problems with the special relativity

Consider an entanglement of two particles 1 and 2 entangled by the paths on which they travel

$$|\psi\rangle = (|a\rangle|b\rangle + |c\rangle|d\rangle)/\sqrt{2}. \tag{1}$$

On the four paths $a$, $b$, $c$ and $d$, are placed non-absorbing detectors.
Consider a trial of the experiment in which at the time $t_1$ by the clock of the lab – the frame S – the particle 1 is tested and found on the path $a$. Therefore, for any $t > t_1$ the particle 2 will travel on the path $b$. Indeed, at a time $t_2 > t_1$ at which the particle 2 is detected it is found on the path $b$. The w-f of the pair of particles between the two events of detection is reduced to $|a\rangle|b\rangle$.

Let now S' be a frame of coordinates in movement with respect to the lab and a clock C' which moves together with the frame S'. The movement of this frame is such that the detection of the particle 2 occurs first, at a time $t'_2$ according to C'. The detection of the particle 1 occurs at $t'_1$, with $t'_1 > t'_2$.

Let's notice that according to S' no detection is performed before $t'_2$, therefore the w-f of the pair of particles before the detection of the particle 2 at must be (1). But returning to the frame S, after the detection of particle 1 and before the detection of particle 2, the w-f is $|a\rangle|b\rangle$.

The conclusion is that the w-f is ambiguously defined; therefore it cannot be ontic, i.e. a real wave traveling through our apparatus.

## 3. The wave-function is not self-interacting

Eliminating the possibility that the w-f is a physical reality, it remains that the entity traveling in our apparatus is a particle. Is was proved in [38] that if the w-f of an electrically charged particle consists in several wave-packets (w-ps), the charge has to be present in each one of the w-ps. Then, the w-ps should repel one another. But no experiment proved such an effect of repulsion. For explaining the absence of such an effect, S. Gao proposed the idea that the particle jumps from w-p to w-p, [37].

However, there is no need to invent jumps for explaining the lack of self-interaction of the w-f. The quantum world behaves differently than the classical rules dictate. Let the w-f of an electron consist in two w-ps $|\psi\rangle_A$ and $|\psi\rangle_B$. According to the second quantization one can write it as

$$|\psi\rangle = (\psi_A |1\rangle_A |0\rangle_B + \psi_B |0\rangle_A |1\rangle_B)/\sqrt{2}, \qquad (2)$$

The quantum formalism distinguishes, as well as the classical physics, between two modes of association of events:
- Logical AND between two events A and B. That means that they occur together. In the w-f (2), the w-p $|1\rangle_A$ does not occur together with the w-p $|1\rangle_B$, but together with the w-p $|0\rangle_B$. If the w-p $\psi_A$ is populated with the charge, the electric force of the charge cannot repel the w-p $\psi_B$ because $\psi_B$ is empty, $|0\rangle_B$. Analogously if $\psi_B$ is occupied. Therefore, $\psi_A$ and $\psi_B$ cannot interact.
- Logical OR means that the events A and B are alternative: either A occurs, or B. In the w-f (2) the presence of the electron in the w-ps $|1\rangle_A$ and $|1\rangle_B$, are alternative situations. Either $|1\rangle_A$ occurs, or $|1\rangle_B$, but they do not occur together.

These rules are the same as in the classical physics with one great difference: in the classical physics $\psi_A |1\rangle_A$ and $\psi_B |1\rangle_B$ do not produce interference.

## 4. The RDM interpretation of QM vs. delayed tests and vs. the special relativity

In his work [35] Gao discusses a couple of difficulties of the RDM interpretation.
These difficulties and additional ones are discussed in this section. For some of the difficulties a solution is proposed.

In the section 7.2.3 of [35] Gao considered two particles, 1 and 2, entangled by position such that when the particle 1 occupies the position $\mathbf{r}_1 = (x_1, y_1, z_1)$, the particle 2 occupies the position $\mathbf{r}_2 = (x_2, y_2, z_2)$, and when the particle 1 occupies the position $\mathbf{r}_3 = (x_3, y_3, z_3)$, the particle 2 occupies the position $\mathbf{r}_4 = (x_4, y_4, z_4)$. Gao assumes that at some time the particle 1 is at $\mathbf{r}_1$ and particle 2 at $\mathbf{r}_2$, and when particle 1 jumps from $\mathbf{r}_1$ to $\mathbf{r}_3$, exactly at the same time the particle 2 jumps from $\mathbf{r}_2$ to $\mathbf{r}_4$.
However, the experimental practice shows that an entanglement remains true also when the two particles are tested at different times. Let particle 1 be tested at $t_1$, exactly after it jumped from $\mathbf{r}_1$ to $\mathbf{r}_3$. Therefore it would be found at $\mathbf{r}_3$. According to Gao's conjecture, the particle 2 jumps at this time from $\mathbf{r}_2$ to $\mathbf{r}_4$. At a later time $t_2$ the particle 2 is tested, when it jumps back to $\mathbf{r}_2$. If the particle 1 were not detected already and found at $\mathbf{r}_3$, it would have jumped back to $\mathbf{r}_1$. But that does not happen anymore.
Bottom line, according to Gao's conjecture, if the particles are detected at different times, the entanglement is violated.
For solving this problem a mechanism should be proposed by which the detection of one particle induces freezing of the other particle in the situation required by the entanglement. Though, from the point of a moving observer for which the particle 2 is tested first, the freezing mechanism must act backwards in time.

Another problem, described by Gao in the section 9.1.1 of [35], is the duplication of the particle. Consider a w-f consisting of two w-ps $\psi_A$ and $\psi_B$ distant from one another. Let the particle be at a time $t_1$ at the point $x_1$ of $\psi_A$ according to a frame of coordinates S, and consider that it jumps at a time $t_2 > t_1$ to the point $x_2$ of $\psi_B$. According to another frame of coordinates, S', moving with respect to S with the velocity $v$, the times and space coordinates are modified as follows

$$t_1' = (t_1 - x_1 v/c^2)/\sqrt{1 - v^2/c^2}, \qquad t_2' = (t_2 - x_2 v/c^2)/\sqrt{1 - v^2/c^2}, \qquad (3)$$

$$x_1' = (x_1 - t_1 v)/\sqrt{1 - v^2/c^2}, \qquad x_2' = (x_2 - t_2 v)/\sqrt{1 - v^2/c^2}, \qquad (4)$$

Setting $t_2' = t_1'$ one gets

$$v = \frac{(t_2 - t_1)c^2}{x_2 - x_1}. \qquad (5)$$

If the jump is nonlocal, i.e. the distance between the two w-ps and therefore between $x_1$ and $x_2$, is greater than $(t_2 - t_1)c$, there results $v < c$. Therefore there exists a frame S' by which the particle is simultaneously in the two w-ps.

This situation cannot be detected experimentally. If an absorbing detector is placed on the way of the w-p $\psi_A$ containing the position $x_1$, it absorbs the particle at $t_1$, so the particle cannot jump anymore to the w-p $\psi_B$.

If the detector is not absorbing, that means that the particle is entangled with another particle, let's name it "ancilla", which has two states, a state $|\xi_1\rangle$ for the main particle being at $x_1$, and a state $|\xi_2\rangle$ for the main particle being at $x_2$. If the detector absorbs the ancilla in the state $|\xi_1\rangle$ at the time $t_1$, the ancilla cannot be re-detected by another detector at the time $t_2$ in a state $|\xi_2\rangle$.

The complementary problem is described by Gao in section 9.1.3 of his work [35]. Consider the w-f $|\psi\rangle = (|\psi_A\rangle + |\psi_B\rangle)/\sqrt{2}$. Let S be the lab frame of coordinates, according to which the two w-ps are tested simultaneously at the time $t_1$. Consider a trial of the experiment in which the w-f collapses on $|\psi_A\rangle$, therefore $|\psi_B\rangle$ vanishes at the same time. Let S' be a frame according to which $|\psi_A\rangle$ meets a detector at the time $t'_1$ and $|\psi_B\rangle$ meets a detector at the time $t'_2$, with $t'_1 > t'_2$. That means, in the interval of time $t'_1 - t'_2$ the detector on $|\psi_B\rangle$ remained silent, and the detector on $|\psi_A\rangle$ wasn't met yet. What remained from the w-f $|\psi\rangle$ is the part $|\psi_A\rangle/\sqrt{2}$, which means that the particle is in $|\psi_A\rangle$ with the probability ½, while with another probability ½ it jumps backward in time, i.e. before $t'_2$, to $|\psi_B\rangle$. In short, the particle is present in $|\psi_A\rangle$ only half of the interval $t'_1 - t'_2$, and during the other half the particle is absent from the space.

This problem too is not observable, what is observable is only what the detectors report. Besides that, the work of the team of S. Savasta, [40], showed that the energy/number-of-particles should be conserved in the states that are detectable, but not in intermediate states (see also [41]).

In the section 9.1.2 if [35] Gao turns to entanglements and describes a more complex problem. Let the entanglement of the particles A and B be prepared as $(|\psi_u\rangle|\varphi_u\rangle + |\psi_d\rangle|\varphi_d\rangle)/\sqrt{2}$ where $|\psi_u\rangle$, $|\psi_d\rangle$ are states of the particle A, and $|\varphi_u\rangle$, $|\varphi_d\rangle$ are states of the particle B. The four w-ps are supposed to travel in regions distanced from one another two by two. Let $t_{1,A}$ be the time at which the particle A jumps from $|\psi_u\rangle$ and let $t_{2,A}$ be the time at which the jump is completed and particle A lands in the w-p $|\psi_d\rangle$, Let $t_{1,B}$ be the time at which the particle B leaves the w-p $|\varphi_u\rangle$ and let $t_{2,B}$ be the time at which it lands in the w-p $|\varphi_d\rangle$. The times $t_{1,A}$, $t_{1,B}$, $t_{2,A}$ and $t_{2,B}$ are readings of the clock in the lab frame of coordinates, S, with $t_{2,A} > t_{1,A}$, $t_{2,B} > t_{1,B}$. Let also $t_{2,B} > t_{1,A}$. Then one can find a frame of coordinates S' in movement with respect to S with the velocity

$$v = \frac{(t_{2,B} - t_{1,A})c^2}{x_{2,B} - x_{1,A}}. \tag{6}$$

In this frame $t'_{2,B} = t'_{1,A}$, with

$$t'_{2,B} = \frac{x_{2,B} t_{1,A} - t_{2,B} x_{1,A}}{(x_{2,B} - x_{1,A})\sqrt{1 - v^2/c^2}}. \tag{7}$$

The equality $t'_{2,B} = t'_{1,A}$ means that by the time the particle B lands in the w-p $|\varphi_d\rangle$, the particle A is in $|\psi_u\rangle$ and is leaving this w-p.

This is a violation of the entanglement.

However, neither this effect could be observed experimentally if Gao would conjecture the existence of a frame S by which once one particle is detected, the other particle stops jumping. If particle A is tested first by this frame and found in $|\psi_u\rangle$ ($|\psi_d\rangle$) the particle B is frozen inside $|\varphi_u\rangle$ ($|\varphi_d\rangle$) until is detected. Then, there are no $t_{2,B}$ and $x_{2,B}$ and the equation (6) has no meaning. By any other frame, no matter what is the order of detection of the particles, the results of the detections will be the same as in S.

A more severe problem, that yes is experimentally observed, is separately presented in the next section.

## 5. A variant of Hardy's paradox

Let's consider a pair of particles, $p^+$ and $p^-$, entangled by the path of flight – figure 1. The w-f is

$$|\psi\rangle = (\imath|\mathbf{u}^+\rangle|\mathbf{v}^-\rangle + |\mathbf{v}^+\rangle|\mathbf{v}^-\rangle + \imath|\mathbf{v}^+\rangle|\mathbf{u}^-\rangle)/\sqrt{3}.^{1} \tag{8}$$

This experiment was proposed and analyzed for the first time by L. Hardy [43], and is known in the literature under the name "Hardy's paradox" – see also [44]. Here it is presented in a modified form which allows testing the QM predictions by direct measurements.

At the beam-splitters $\mathbf{BS}^{\pm}$ occur the transformations

---

[1] This w-f was prepared experimentally by Lundeen et al. [42].

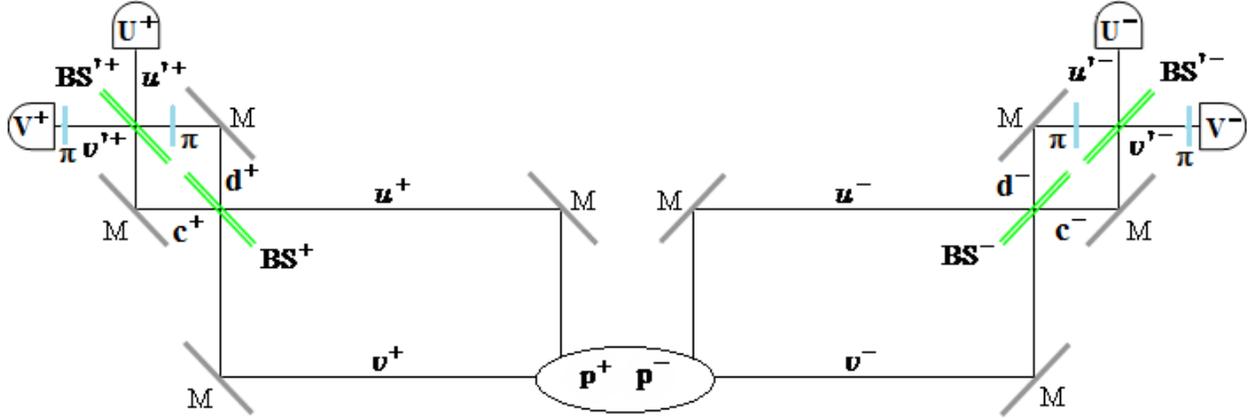

Figure 1. An experiment with a two-particle entanglement.
$\mathbf{u}^\pm$, $\mathbf{v}^\pm$, $\mathbf{c}^\pm$, $\mathbf{d}^\pm$, $\mathbf{u'}^\pm$, $\mathbf{v'}^\pm$, are directions of flight of the particles. $\mathbf{BS}^\pm$ and $\mathbf{BS'}^\pm$ are beam-splitters. M are mirrors. $U^\pm$ and $V^\pm$ are detectors. The region $p^+p^-$ is the preparation region of the entanglement. $\pi$ symbolize phase-shifters by $\pi$.

$$|\mathbf{u}^\pm\rangle \to (|\mathbf{c}^\pm\rangle + \iota|\mathbf{d}^\pm\rangle)/\sqrt{2}, \quad |\mathbf{v}^\pm\rangle \to (\iota|\mathbf{c}^\pm\rangle + |\mathbf{d}^\pm\rangle)/\sqrt{2}. \tag{9}$$

Introducing them in (8) one gets

$$|\phi\rangle = (-3|\mathbf{c}^+\rangle|\mathbf{c}^-\rangle + \iota|\mathbf{c}^+\rangle|\mathbf{d}^-\rangle + \iota|\mathbf{d}^+\rangle|\mathbf{c}^-\rangle - |\mathbf{d}^+\rangle|\mathbf{d}^-\rangle)/\sqrt{12}. \tag{10}$$

The relevant term in this w-f is $|\mathbf{d}^+\rangle|\mathbf{d}^-\rangle$ implying that if the detectors were moved to the paths $\mathbf{c}^+$, $\mathbf{d}^+$, $\mathbf{c}^-$, $\mathbf{d}^-$, the detectors $D^+$ and $D^-$ would click simultaneously. The probability of obtaining this joint detection would be 1/12. Returning to the figure 1 with the detectors on the paths $\mathbf{u}^\pm$ and $\mathbf{v}^\pm$, by the RDM interpretation, prior to meeting the beam-splitters $BS'^\pm$ the two particles were on the paths $\mathbf{d}^+$ and $\mathbf{d}^-$ respectively, with the probability 1/12.

The transformations at the beam-splitters $BS'^\pm$ and taking also in consideration the phase-shifts, are

$$|\mathbf{c}^\pm\rangle \to (|\mathbf{u'}^\pm\rangle - \iota|\mathbf{v'}^\pm\rangle)/\sqrt{2}, \quad |\mathbf{d}^\pm\rangle \to -(\iota|\mathbf{u'}^\pm\rangle - |\mathbf{v'}^\pm\rangle)/\sqrt{2}. \tag{11}$$

Introducing them in (10) one gets back the initial w-f (8):

$$|\psi'\rangle = \left(\iota|\mathbf{u'}^+\rangle|\mathbf{v'}^-\rangle + \iota|\mathbf{v'}^+\rangle|\mathbf{u'}^-\rangle + |\mathbf{v'}^+\rangle|\mathbf{v'}^-\rangle\right)/\sqrt{3}. \tag{12}$$

Let's now imagine that the setup in the figure 1 is sufficiently wide so that one can find a frame of coordinates S' flying in the direction from BS'$^-$ to BS'$^+$, on whose time axis, by the time the detector $U^+$ clicks, the w-ps of $p^-$ didn't reach yet $BS'^-$. Introducing in (10) the transformations (11) only for the particle $p^+$, one gets

$$|\phi^+\rangle = (2\iota|\mathbf{v}'^+\rangle|\mathbf{c}^-\rangle + |\mathbf{u}'^+\rangle|\mathbf{c}^-\rangle + \iota|\mathbf{u}'^+\rangle|\mathbf{d}^-\rangle)/\sqrt{6}. \tag{13}$$

One can see that by the time, on the time axis of S', that while the particle $p^-$ travels along the path $d^-$, $p^+$ must click the detector $U^+$.

But one can also find a frame of coordinates S" flying in the direction from $BS'^+$ to $BS'^-$, on whose time axis, by the time the detector $U^-$ clicks, the w-ps of $p^+$ didn't reach yet $BS'^+$. Introducing in (10) the transformations (11) only for the particle $p^-$, one gets

$$|\phi^-\rangle = (2\iota|\mathbf{c}^+\rangle|\mathbf{v}'^-\rangle + |\mathbf{c}^+\rangle|\mathbf{u}'^-\rangle + \iota|\mathbf{d}^+\rangle|\mathbf{u}'^-\rangle)/\sqrt{6}. \tag{14}$$

That implies that by the time, according to S", that while the particle $p^+$ travels along the path $d^+$, $p^-$ must click the detector $U^-$.

Returning now to the frame of coordinates of the lab with respect to which the source, the beam-splitters and the mirrors are at rest, the w-f of the two particles is (12). And it can't allow both detectors $U^+$ and $U^-$ click together because it contains no combination $|\mathbf{u}'^+\rangle|\mathbf{u}'^-\rangle$.

According to the RDM interpretation the two particles should travel simultaneously on the paths $d^+$ and $d^-$, but if the requirements of the special relativity are taken in consideration, one comes to a contradiction.

## 6. Conclusions

The most wide-spread interpretations of the QM were examined here and it was shown that they present inconsistencies. Along with this, it was shown that the idea that the w-f is a real wave, has problems with the special relativity. One would then turn to the option that the w-f is just a tool for calculating probabilities of results of measurements, and what is real in the QOs are particles. Since it was proved that the particles can't follow continuous trajectories, they should jump at random from place to place. Thus, one is lead to Gao's RDM interpretation of the QM. It was though proved here that neither this interpretation can overcome the difficulties with the special relativity. Thus, it appears that none of the most wide-spread interpretations of the QM is free of problems.